# Dynamics of transmission in disordered topological insulators

Yuhao Kang, Yiming Huang and Azriel Z. Genack[*]

Queens College and The Graduate Center of the City University of New York

Flushing, NY 11367, USA

Robust transmission in topological insulators makes it possible to steer waves without attenuation along bent paths within imperfectly fabricated photonic devices. But the absence of reflection does not guarantee the fidelity of pulsed transmission which is essential for core photonic functionalities. Pulse transmission is disrupted by localized modes in the bulk of topological insulators which coexist with the continuum edge mode and are pushed deeper into the band gap with increasing disorder. Here we show in simulations of the Haldane model that pulse propagation in disordered topological insulators is robust throughout the central portion of the band gap where localized modes do not arise. Since transmission is robust in topological insulators, the essential field variable is the phase of the transmitted field, or, equivalently, its spectral derivative, which is the transmission time. Except near resonances with bulk localized modes that couple the upper and lower edges of a topological insulator, the transmission time in a topological insulator is proportional to the density of states and to the energy excited within the sample. The variance of the transmission time at the band edge for a random ensemble with moderate disorder is dominated by fluctuations at resonances with localized states, and initially scales quadratically. When modes are absent, such as in the center of the band gap, the transmission time self-averages and its variance scales linearly. This leads to significant sample-to-sample fluctuations in the transmission time. However, because the transmission time is the sum of contributions from the continuum edge mode, which stretches across the band gap, and far-off-resonance modes of the passband and band edge, there are no sharp features in the spectrum of transmission time in the center of the band gap. Instead of creating obstacles to be circumvented with inevitable increase of the transmission time in the band edge mode, the transmission time of the band edge mode is decreased by disorder. As a result, ultrashort, broadband pulses are faithfully transmitted in the center of the band gap of topological insulators with moderate disorder and bent paths. This allows for robust signal propagation in complex topological metawaveguides for applications in high-speed optoelectronics and telecommunications.

Subject Areas:

[*]genack@qc.edu

# I. INTRODUCTION

There has been rapid progress in understanding and implementing photonic topological insulators (TIs) in microwave and optical metawaveguides and metamaterials since the electromagnetic analogy to topologically protected chiral edge states in various quantum Hall effects was first proposed a decade ago [1–12]. Robust unidirectional propagation across the band gap of the bulk of a 2D material along the boundary line between regions with different topological invariants opens the way towards novel applications and compact device architectures. These include new possibilities for optical isolators, and for optical buffering and signal processing on a chip [10,13,14]. Most potential applications in photonic TIs depend not only on the robustness of transmission but also on the fidelity of pulsed transmission in the presence of nanoscale disorder that occur in fabricated metawaveguides. Robust chiral edge states have been demonstrated for acoustic waves in anomalous Floquet topological insulators in strongly coupled ring lattices [15] in which pulses travelling along the edge are delayed by square-shaped detours with four sharp bends [16]. However, the statistics of dynamics in homogeneously disordered Tis has not been studied.

Since flux is perfectly transmitted through a TI, it is only the phase of the transmitted field that can vary as the frequency is tuned. The spectral derivative of the phase is equal to the transmission time, $\frac{d\varphi}{d\omega} = \tau_T$ [17–27]. In a pioneering study, Mittal *et al.* studied the transmission time through the edge states of 2D lattices with a synthetic gauge field with sizes of up to $15 \times 15$ coupled silicon ring resonators [7]. They found in samples with intrinsic fabrication imperfections that the distribution of the optical transmission time normalized to its average across the band gap, $P(\tau_T/\langle \tau_T \rangle)$, is approximately Gaussian with width independent of sample size. This corresponds to quadratic scaling of the variance of the transmission time, $\text{var}(\tau_T)$. Based on the analogy with the statistics of transmission time in 1D coupled-resonator optical waveguides (CROWs) with silicon resonators [28], they suggested that transport is diffusive in edge states even when disorder is strong enough that waves are localized in the bulk [7].

But a closer examination is necessary because there is no diffusive regime in random 1D systems since the localization length is equal to the mean free path, $\xi = \ell$; waves are either ballistic, $L < \ell$, or localized, $L > \xi$. In addition, we show below that $\text{var}(\tau_T)$ in random 1D media scales nearly quartically rather than quadratically. Given the changing density of localized bulk modes across the band gap [29–31], it is valuable to explore the possibility of distinctive statistics of transmission time near the edges and the center region of the band gap.

The dynamics of transmission in nontopological random media is linked to the spatial extent of the quasi-normal modes of the system. Modes extend throughout the sample in diffusive systems and fall exponentially towards the sample boundaries when waves are localized [32,33]. Because localized modes are weakly coupled to the boundaries, they are long-lived and exhibit sharp spectral peaks in transmission [33,34]. However, since transmission in a TI is robust, modes are not manifest in peaks in the transmission spectrum. The nature of transport and the states underlying wave propagation can be uncovered, however, in spectra of transmission time and the area speckle pattern within the medium.

Here, we explore the dynamics of propagation in disordered TIs in simulations of the tight-binding Haldane model [35]. This provides a basis for evaluating the viability of TI metawaveguides fabricated with inevitable disorder for routing pulses along bent paths [12,36,37]. Localized modes in the band gap of disordered TI differ from those in trivial random media since they coexist with the extended continuum edge state and their spectral width reflects their distance from the edge, not from the input or output boundaries of the sample. Localized modes appear near the band edge for moderate disorder and move deeper into the band gap with increasing disorder, until the band gap is washed out for extreme disorder [29–31,38,39]. The transmission time, $\tau_T$, is proportional to the density of states (DOS), $\rho$, and to the energy excited inside the medium for unit incident flux in all channels connecting the medium to its environment, $U$, when systems are either perfectly reciprocal or perfectly nonreciprocal. Large fluctuations in $\tau_T$ arise near resonance with localized states near the band edge. But fluctuations in the DOS also arise in the center of the band gap where localized modes are not present. These fluctuations are associated with the continuum edge mode and the tails of modes of the passband and band edge, which are perturbed by disorder. These factors produce significant sample-to-sample fluctuations at the center of the band gap, but to a smooth spectrum of $\tau_T$ in any single realization of disorder. Thus, fluctuations in $\tau_T$ relative to its value in the pristine TI arise throughout the band gap with properties that depend upon the strength of disorder and the frequency shift from the band edge. For weak disorder near the band edge or moderate disorder near the center of the band gap, modes do not form and $\tau_T$ self-averages and $\text{var}(\tau_T)$ scales linearly. For moderate disorder near the band edge, however, $\text{var}(\tau_T)$ initially scales quadratically. This scaling differs from the nearly quartic universal scaling of $\text{var}(\tau_T)$ in trivial 1D random media. The essential point for applications is that disordered TIs are thus nonergodic near the band center in the sense that the range of possible values of $\tau_T$ in an individual sample is smaller than the range in the ensemble. Because, the dispersion in $\tau_T$ near its minimum in the center of the band gap in each sample is small, however, broadband pulses can be transmitted with minimal distortion. The impact of dispersion can be addressed using approaches that are employed fiber optic communications. Thus, disorder does not limit the fidelity of pulse propagation which is crucial for applications in signal processing, isolation, and communication.

## II. RESULTS AND DISCUSSION

### A. Haldane model

We simulate propagation in a honeycomb lattice in the Haldane model [35] with the Hamiltonian

$$H = \sum_i c_i^\dagger \eta_i c_i + \sum_{\langle ij \rangle} c_i^\dagger t c_j + \sum_{\langle\langle ij \rangle\rangle} c_i^\dagger i \lambda_{SO} v_{ij} c_j. \tag{1}$$

Here $c_i^\dagger (c_i)$ denotes creation (annihilation) operators at site $i$, the onsite energy $\eta_i$ is uniformly distributed between $-W/2$ and $W/2$, and $t=1$ is the uniform hopping rate between nearest neighbors $\langle ij \rangle$. $\lambda_{SO} = 0.1$ is the spin-orbit coupling strength, and $v_{ij} = (\hat{d}_1 \times \hat{d}_2)/|\hat{d}_1 \times \hat{d}_2|$, where $\hat{d}_1$ and $\hat{d}_2$ are two nearest neighbor bonds connecting next-nearest-neighbor $\langle\langle ij \rangle\rangle$. The lattice

constant is 1. The scattering region is attached to two semi-infinite ordered leads, as shown in Fig. 1a so there is only a single mode entering and leaving the sample, with the upper (lower) boundary supports a right- (left-) moving edge state. Simulations are carried out using the open-source package Kwant [40].

### B. Propagation at the band edge

We first consider propagation near the band edge where localized modes in the bulk coexist with the extended edge state. The center of the band gap is at $\omega = 0$. We obtain the transmitted signal in the right lead of a TI of finite width for a wave injected into the left lead. The Green's function between the input and output at the upper boundary in a typical sample is shown in Fig. 1b with the magnitude and phase of the field given by the blue and red curves, respectively. The magnitude is a constant baseline with several dips, while the phase shows regions of nearly constant slope between sharp jumps of $2\pi$ as the wave is tuned through resonance with modes in the bulk. Drops in transmission occur when the wave in the upper edge tunnels through a localized mode in the bulk to the lower edge. The wave amplitude in the interior of the sample at the frequencies marked c and e in Fig. 1b around which the phase of the Green's function increases at a nearly constant rate can be seen in Figs. 1c,e to be similar. The excitation in these regions is due to the continuum edge state and modes in the passbands which change slowly with frequency. The peaks circled in Figs. 1c,e are spectrally broad because they are strongly coupled to the upper edge. Figure 1d shows a sharply peaked localized mode further from the upper edge at the peak of the resonance indicated by d in Fig. 1b.

`

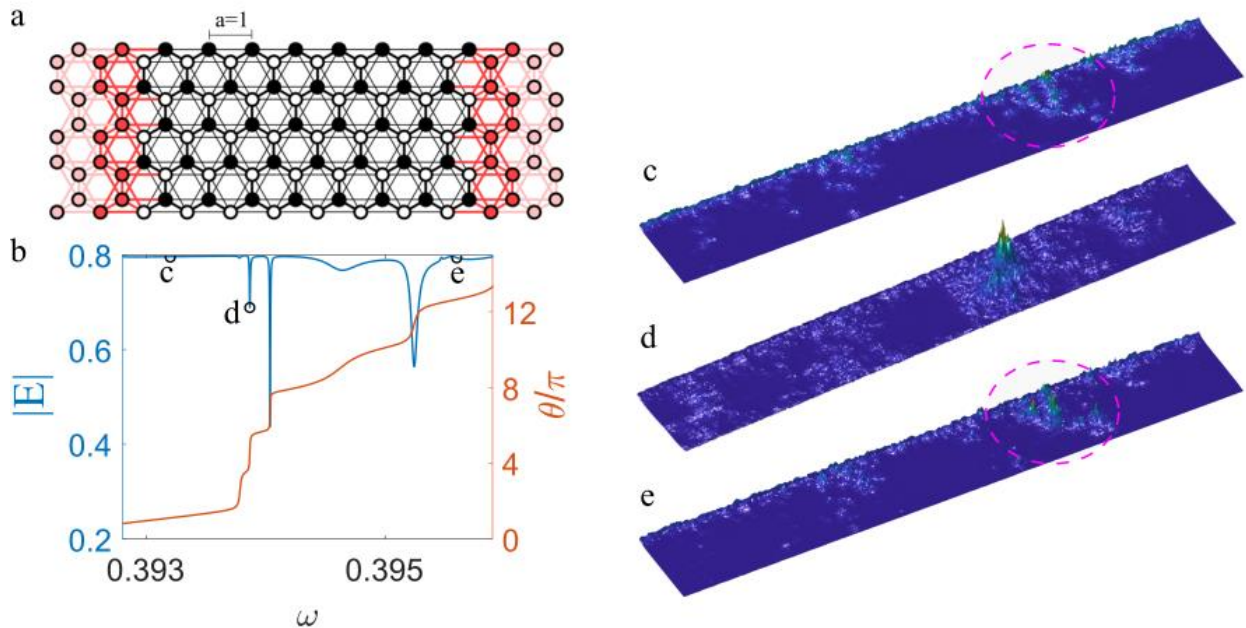

FIG. 1 (color online). Numerical simulation of propagation in the disordered Haldane model near the band edge. (a) The disordered region is indicated in black-white, while the pink region at the ends of the sample represent the pristine leads. The lattice constant a is 1, the sample has width, length, and disorder strength of $w=60$, $L=400$, and $W=1.6$, respectively. (b) The transmitted field through the upper edge. The blue and red curves give the magnitude and phase of the transmitted field, respectively. (c-e) The spatial distribution of intensity at the three frequencies indicated in (b).

The transmitted field can be expressed as the superposition of the continuum edge mode, $E_0$, and the contributions of quasi-normal modes, $G(r',r,\omega) = \langle r'|\dfrac{1}{\omega - H_{eff}}|r\rangle = \sum_n \dfrac{\langle r'|\psi_{nr}\rangle\langle\psi_{nl}|r\rangle}{\omega - \omega_n + i\Gamma_n/2}$, where $|\psi_{nr}\rangle$ and $\langle\psi_{nl}|$ represent the biorthogonal basis of $H_{eff}$, and $\omega_n$ and $\Gamma_n$ are the central frequency and linewidth of the $n^{th}$ discrete mode. Thus, the field at the output can be expressed as

$$E = E_0 + \sum_n \dfrac{V_n}{\omega - \omega_n + i\Gamma_n/2}, \qquad (2)$$

where $V_n$ is the amplitude coefficient of the $n^{th}$ discrete mode [41]. In practice it is not possible to distinguish the featureless contributions of the off-resonance contributions of modes in the passband and the continuum edge mode to the spectrum of the transmitted field. These contributions will therefore be grouped together and denoted as the field due to the continuum, $E_0$, while the modal sum will represent only modes with discernable resonances in the spectrum which is analyzed.

Contributions of the continuum and localized modes to simulations of the spectrum of the field transmitted through the sample shown in Fig. 1a are found by fitting Eq. (2) to the field spectrum. This gives the magnitude and phase of $E_0$, which are plotted as the thin blue and red curves in Fig. 2a. In order to find the resonant contributions to the field, we first determine the phase of $E_0$, $\theta$, by a linear fit to the spectrum of phase in a region between sharp modes. The transmission time associated with the continuum in the disordered TI at $\omega = 0.394$, $d\theta/d\omega = 744$, is smaller than the transmission time in the periodic TI of $t_0 = 967$. This may be due to the removal of the contribution of discrete modes near the band edge to $\tau_T$ in the disordered system using Eq. (2).

The magnitude of $E_0$ can be obtained through an interpolation of the baseline shown in Fig. 2a. The separation of the field into continuum and resonance contributions requires a model of modal excitation for the proper analysis of the variation of the phase near resonance. The increase of phase by $2\pi$ when tuning through a resonance in a disordered TI seen in Figs. 1b,2a contrasts with the increase of $\pi$ in a trivial random medium. The phase change of $2\pi$ can be understood from the complex representation of the field shown in Fig. 2b in which the curve of the complex field encircles the original point in the presence of the continuum. Subtracting the interpolated continuum field $E_0$ from the total field gives the transmitted field due to resonances, as shown in Fig. 2c. This spectrum is fit using the method of harmonic inversion [42] to give the contribution of individual modes, $E_n = \dfrac{V_n}{\omega - \omega_n + i\Gamma_n/2}$, which is plotted in Fig. 2d.

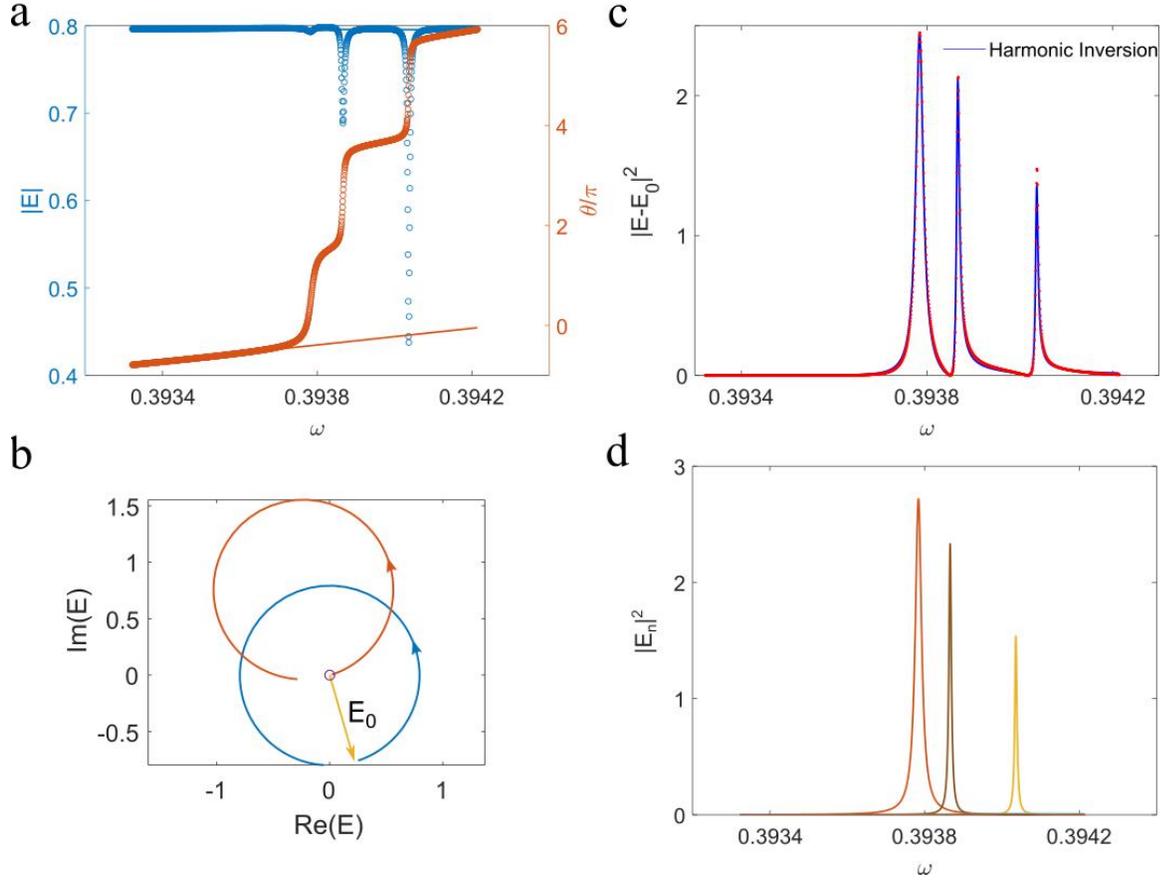

FIG. 2 (color online). Decomposition of the field into the sum of a continuum and discrete resonances. (a) A section of the spectrum shown in Fig. 1b. The thin lines correspond to the contributions of the continuum $E_0$ obtained from a linear extrapolation. (b) The complex representation of the first resonant mode seen in (d) (red), the edge mode (yellow arrow) and the sum, which is the total field (blue). (c) The dotted red curve is the intensity associated with $E-E_0$. The blue curve is determined by a modal fit. (d) The Lorentzian functions for each mode are $|E_n|^2 = \frac{|V_n|^2}{(\omega-\omega_n)^2+(\Gamma_n/2)^2}$.

The analysis of the field using Eq. (2) reveals the nature of the phase variation of the continuum field and modes in TIs, however, its application is limited to narrow frequency ranges in which the slope of $\theta$ is nearly constant. We therefore undertake a more general approach to modal decomposition, which also illuminates the way energy is excited within disordered systems. In lossless reciprocal media, the DOS is proportional to the Wigner time delay, $\tau_W$, and to the energy integral over the scattering region when all incoming channels are excited by unit flux, $U$, $\rho = \frac{1}{2\pi}\tau_W = \frac{1}{2\pi}U$ [27,43,44]. Here, $\tau_W = -iTr\left(S^\dagger \frac{dS}{d\omega}\right)$, where the scattering matrix $S$ gives the field transmission coefficients between all channels [19,25,26]. In reciprocal systems, the transmission time is the same for a wave launched from either side of the sample and the Wigner time is twice $\tau_T$, $\tau_W = 2\tau_T$ [27], so that $\tau_T = \pi\rho = U/2$. The last equality is demonstrated for a reciprocal system in section I of the Supplementary Materials [45]. In a perfectly nonreciprocal TI, $\tau_W = \tau_T$ giving

$$\tau_T = 2\pi\rho = U. \tag{3}$$

Here, the U indicates the intensity integral when the upper boundary is excited by the mode propagating to the right. In Fig. 3a, we compare $\tau_T$ with $U$ in a sample with W=1.6, w=60, and L=200. Differences between $\tau_T$ and $U$ are seen in the dip in Figs. 3a,b near $\omega$=0.3965. The speckle pattern within the sample at this frequency in Fig. 3c shows a localized mode within the red circle. This mode excited near the lower boundary couples the wave propagating to the right in the upper edge to the wave propagating to the left in the lower edge of the sample. Figure 3d shows the Lorentzian line in the spectrum of reflection time, $\tau_R = \frac{d \arg(r)}{d\omega}$, where $r$ is the reflected field. When the reflection is not negligible, $\tau_T$ and $U$ are no longer equal, however, the Wigner time $\tau_W$ is still equals to the intensity integral $U$ when all incoming channels are excited, as shown in section II of the SM [45]. When transmission is perfect, the analysis of the central frequency and linewidth of the modes inside the sample is facilitated by Eq. (3) with $\tau_T$ expressed as the sum of a continuum and Lorentzians lines, as shown in the Supplementary Material [45],

$$\tau_T = 2\pi\rho = 2\pi\rho_0 + 2\sum_n \frac{\Gamma_n/2}{(\omega-\omega_n)^2+(\Gamma_n/2)^2}. \tag{4}$$

In principle, $\rho_0$ is the DOS associated with the edge mode, but, in practice, it also includes the contributions of the tails of modes in the passband.

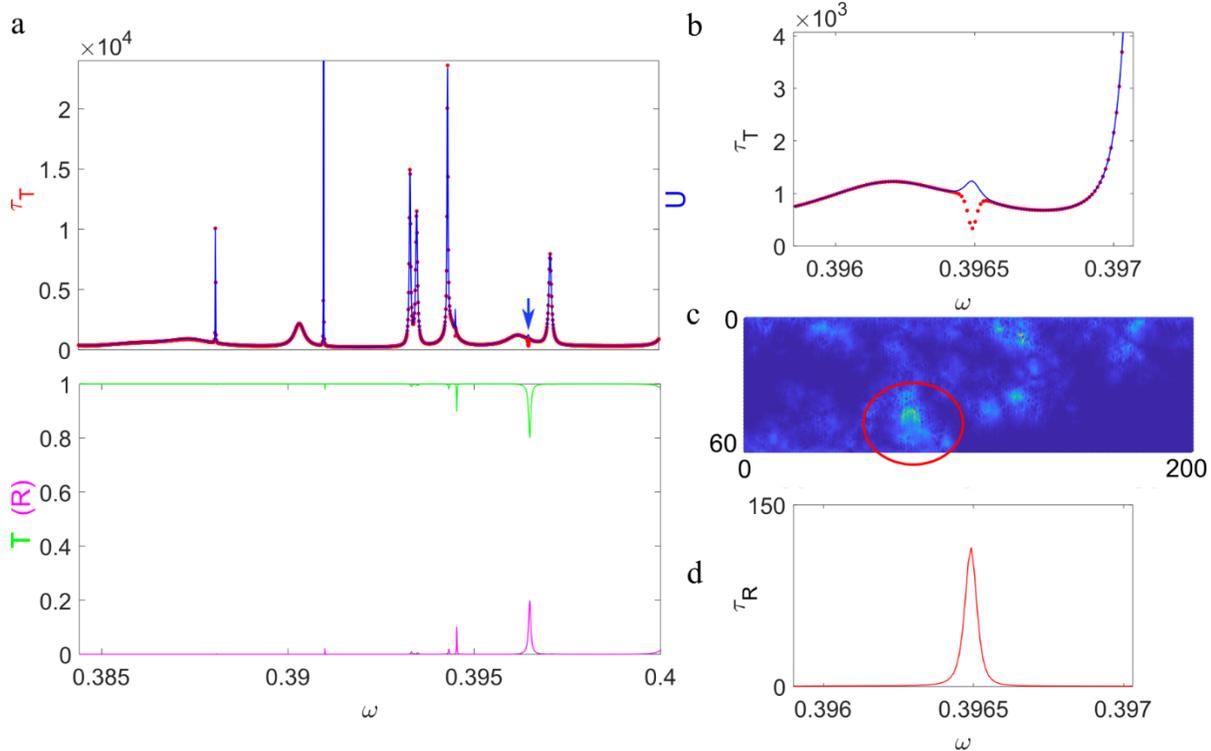

FIG. 3 (color online). Transmission and reflection times in the Haldane model. In this sample, $w=60$, $L=200$, and $W=1.6$. (a) The red and blue curves are $\tau_T$ and $U$, respectively, for a wave injected from the left. The transmission (blue) and reflection (red) spectra are shown in the lower panel. (b) Zoom-in of (a) in the region indicated by an arrow in (a). (c) The intensity pattern at $\omega = 0.3965$. A mode in the region circled in red gives the Lorentzian line in the reflection spectrum shown in (d).

We now consider the scaling and statistics of $\tau_T$ and pulsed transmission near the band edge at $\omega = 0.35$ in TIs with different degrees of disorder. The scaling of the ensemble average of transmission time, $\langle \tau_T \rangle$ normalized by the ballistic time in the pristine TI, $t_0$, is shown in Fig. 4a. In samples with W=0.6 and 1.3, $\langle \tau_T \rangle / t_0$ is nearly independent of length and is elevated above $t_0$ by approximately 1 and 10%, respectively. For stronger disorder, $\tau_T$ is significantly higher than $t_0$ and $\langle \tau_T \rangle / t_0$ increases and appears to approach saturation.

The impact of disorder upon pulsed transmission near the band edge is seen in Fig. 4b. A Gaussian pulse with bandwidth $8 \times 10^{-4}$ and central frequency 0.3937 injected into a sample of length $L=400$ is transmitted without significant distortion for a sample with W=0.6. In this sample, modes are not pushed out of the pass band to this frequency. When the same pulse is injected into the sample with W=1.53 with the spectrum shown in Fig. 3a, the transmitted signal oscillates with a frequency corresponding to the frequency difference between the two modes of the medium that fall within the spectrum of the incident pulse, and decays in the decay rate of the modes excited. These differences in pulse transmission in samples with W=0.6 and 1.53 are accompanied by distinctive differences in the statistics of $\tau_T$ over ensemble of disordered TIs.

Probability distribution functions of the transmission time normalized by its average, $P(\tau_T/\langle\tau_T\rangle)$, for a number of lengths for these two strengths of disorder are plotted in Figs. 4c,d. For W=0.6, $\tau_T$ self-averages. $P(\tau_T/\langle\tau_T\rangle)$ narrows and approaches a Gaussian distribution as the sample length increases and the variance of the transmission time scales linearly, as seen in Figs. 4c,e. This reflects the topological proscription of backscattering which implies that the transit times through different segments of the sample are statistically independent and the net transmission time is the sum of the transit times in the individual segments.

The approach towards a narrowed Gaussian distribution of $\tau_T$ with increasing length and the linear scaling of its variance might also be expected to occur in more strongly scattering samples, where backscattering is similarly prohibited. However, as is foreshadowed in the breakup of an incident pulse in the ensemble with W=1.53 shown in Fig, 4b, the statistics of dynamics changes when disorder increases. At a disorder of W=1.53, $P(\tau_T/\langle\tau_T\rangle)$ appears to converge to a fixed distribution over the length scale studied in the simulations, as seen in Fig. 4d. This corresponds to the quadratic scaling of $\text{var}(\tau_T)$ for the bulk of the distribution, $\tau_T < 10\langle\tau_T\rangle$, seen in Fig. 4f. For this ensemble, the sum of transmission times in the two halves of a sample is equal the total transmission time, as expected. Further, the transmission times of the two halves are not correlated, as is borne out in the scatterplot of transit times in the two halves of a sample shown in section V of the SM [45]. The source of the difference in the statistics for the two samples is that sharp peaks in $\tau_T$ associated with localized modes appear in the spectrum of $\tau_T$ for moderate disorder of W=1.53 but do not appear for W=0.6 in this region of the band edge. These peaks dominate $\text{var}(\tau_T)$ in the sample with moderate disorder. For *L*=500, $\text{var}(\tau_T)$ can be seen in Figs. 4e,f to be larger by a factor of nearly $3 \times 10^3$ for the ensemble with W=1.53 as compared to the sample with W=0.6. The long-tail in the distribution of $\tau_T$ in Fig. 4d is a general feature that arises when the edge mode interacts with bulk modes. In section IV of the SM [45], we measured the transmission time distribution of a disorder photonic system emulating the Kane-Mele model, we also obtain a similar long-tail in the distribution of $\tau_T$.

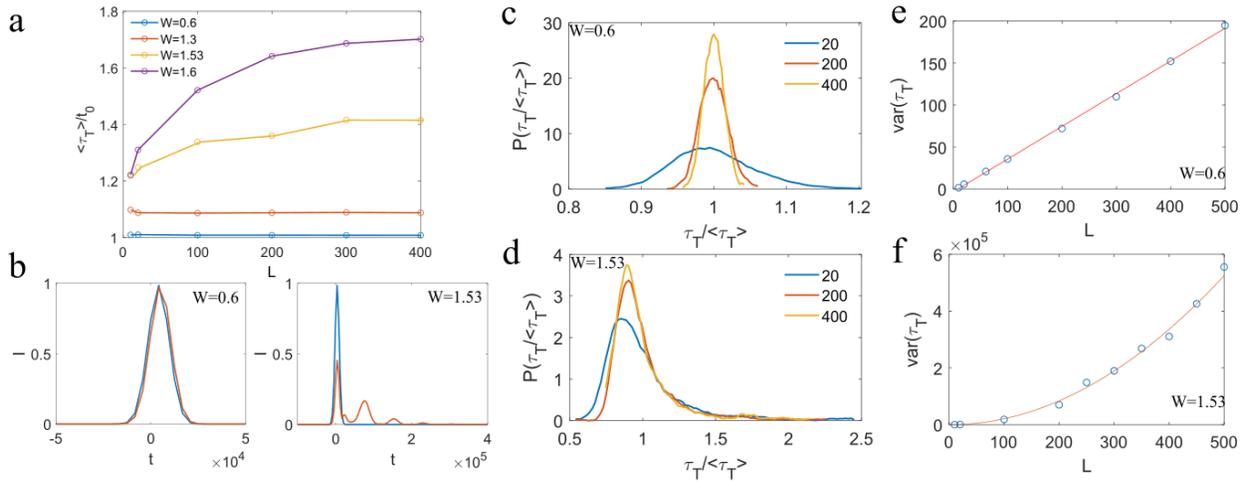

FIG. 4 (color online). Scaling and statistics of transmission time and pulsed transmission for different strengths of disorder near the band edge. (a) The scaling of transmission time relative to the ballistic time in the pristine sample for different disorder strengths in samples with *w*= 60 at $\omega = 0.35$. (b) The blue curves are the input Gaussian pulse with bandwidth $8 \times 10^{-4}$ and central frequency 0.3937, the red curves show the transmitted signal at the output for a sample with W=0.6 (left) and W=1.53 (right). (c,d) The probability distributions of transmission time for $\omega = 0.35$ normalized by the average value for W=0.6 and 1.53, respectively. (e) Scaling of the variance of the transmission time for W=0.6 and the linear fit, giving $\text{var}(\tau_T) = 0.39L$. (f) Scaling of the variance of the bulk of the distribution of transmission time relative to the length for W=1.53 using values of $\tau_T/\langle\tau_T\rangle < 10$ and a quadratic fit giving $\text{var}(\tau_T) = 2.1L^2$.

### C. Propagation in the center of the band gap

Spectra of $\tau_T = 2\pi\rho = U$, (Eq. (3)) across the entire band gap are shown in Fig. 5a for a periodic TI and for three random configurations in samples drawn from the same ensemble with W=1.6 and *L*=400. The absence of peaks in $\tau_T$ in the central portion of the band gap indicates that localized states are not formed away from the band edge and that the upper and lower edges are not coupled. The central portions of the spectra of $\tau_T$ shown in Fig. 5a are displayed in Fig. 5b on a scale that allows these spectra to be compared. The spectrum of $\tau_T$ in different regions of the band gap of a disordered TI are determined by the contribution to the DOS of the continuum edge mode and by the on- and off-resonance contributions of the quasi-normal modes of the bulk material. The relationship between $\tau_T$ and the various contributions to the DOS near the center of the band where transmission is robust is given by Eq. (4).

The spectra of $\tau_T$ for the three disordered TIs in Figs. 5a,b have different minima and frequencies at which these minima occur, but, near the center of the band gap, all the spectra fall below the spectrum of the ordered TI. The spectrum of the transmission time of the edge mode, $t_0$ in the periodic TI is shown as the blue dots in Fig, 5b. Here, $t_0 = L/v_g$, where $v_g = d\omega/dk$ is obtained from the dispersion of the edge mode in the Haldane model shown in the inset of Fig. 5a. The average of the transmission time in the disordered TI at $\omega = 0$, $\langle\tau_T\rangle = 680$, is 5% lower than the values in the periodic TI of $\tau_T = t_0 = 719$. For many samples, the minimum in $\tau_T$ is lower than 680. The overlap of spectra of $\tau_T$ and $t_0$ in the band center for the periodic TI seen in Fig. 5b indicates that the contribution of the continuum edge mode to the DOS overwhelms the contribution of the modes of the passband. The reduction in $\langle\tau_T\rangle$ is therefore indicates that velocity of the edge mode is enhanced by disorder. This is surprising in view of the expected increase in $\tau$ in samples in which direct transmission along the edge is blocked by barriers at points along the edge of a TI.

It is natural that the edge mode will dominate the DOS in the band gap since it stretches across the entire sample with undiminished intensity. But, for the same reason, one might expect the off-resonance contribution of modes in the passband to be significant as well since energy might be excited within the bulk by the energy along the entire length of the sample in the edge mode. In comparison, energy is excited in nontopological nearly-periodic systems near the surface upon which energy is incident. The DOS in the band gap of topologically trivial media is entirely due to modes associated with the passband of the periodic system. The energy density in the band gap falls rapidly away from the incident surface so that the transmission time saturates with sample length, but the DOS does not vanish. Since the minimum transmission in a nontopological band gap for a sample that is periodic on average is for the periodic sample [46], the average decay length of intensity inside the sample is smallest in this case. In accord with this result, we find that $\langle \tau_T \rangle$ in the center of the band gap of nontopological random systems is increased by disorder, as discussed in the Supplemental Material [45]. This is in contrast to the decrease in $\langle \tau_T \rangle$ in topological TIs seen in Fig. 5b. The reduction in $\langle \tau_T \rangle$ in disordered TIs in the center of the band gap also stands in contrast to the enhancement by over 70% in the sample with W=1.6 for $\omega = 0.35$ (Fig. 4a), and to a much greater enhancement in $\langle \tau_T \rangle$ at values of $\omega$ closer to the band edge (Fig. 5a).

Disorder affects both the central frequency and linewidth of the modes of the passband of the periodic system. The central frequencies of modes in the passband of the periodic system are randomized by disorder with the long-lived modes at the band edge perturbed the most. Because modes tend to be pushed into the band gap, the widths of the effective band gaps are reduced. This increases the curvature of $\tau_T$ in the disordered TI, as seen in the inset of Fig. 5b. The linewidths of modes are also change by disorder. Linewidths are generally reduced by disorder since their lifetimes tend to lengthen as modes become more localized. The increase in $\langle \tau_T \rangle$ near the band edge is due to resonances with localized modes. This is seen in the spectra of $\tau_T$ in the individual configurations in Figs. 1-3,5a. In samples with smaller, but still substantial, values of W, modes do not form far from the band edge of the periodic system and the enhancement in $\langle \tau_T \rangle$ close to the band edge is small (Fig. 4a).

Since $\tau_T \sim t_0$ in the pristine TI, the quasi-normal modes of the medium do not contribute substantially to the DOS, but it is possible they do contribute to the DOS of the disordered system. Since the DOS in the center of the band gap is raised on average in topologically trivial systems by the increased contribution of modes in the passband, this should be the case as well in TIs. But since $\langle \tau_T \rangle$ in the disordered TI is lower than in the ordered TI, the drop must be due primarily to the drop in the DOS of the edge mode.

The statistics of $\tau_T$ in an ensemble of 1000 random realizations of the disordered TI with W=1.6 near the center of the band gap is seen in Figs. 5c,d to self-average: the distribution $P(\tau_T/\langle \tau_T \rangle)$ narrows with increasing sample length and approaching a Gaussian function, while $\text{var}(\tau_T)$ scales linearly. This is similar to the statistics of $\tau_T$ near the band edge with disorder of W=0.6 which is not strong enough to introduce modes into the band gap. $\text{var}(\tau_T)$ in a sample with *L*=400 and W=1.6 near the center of the band (Fig. 5d) is smaller than near the band edge for W=0.6 (Fig. 4e), where the contribution of quasi-normal modes is greater. Since the modes of the passband do not appear to influence $\tau_T$ near the band center, fluctuations in $\tau_T$ characterized in Figs. 5c.d are due primarily to fluctuations in the edge mode in different configurations.

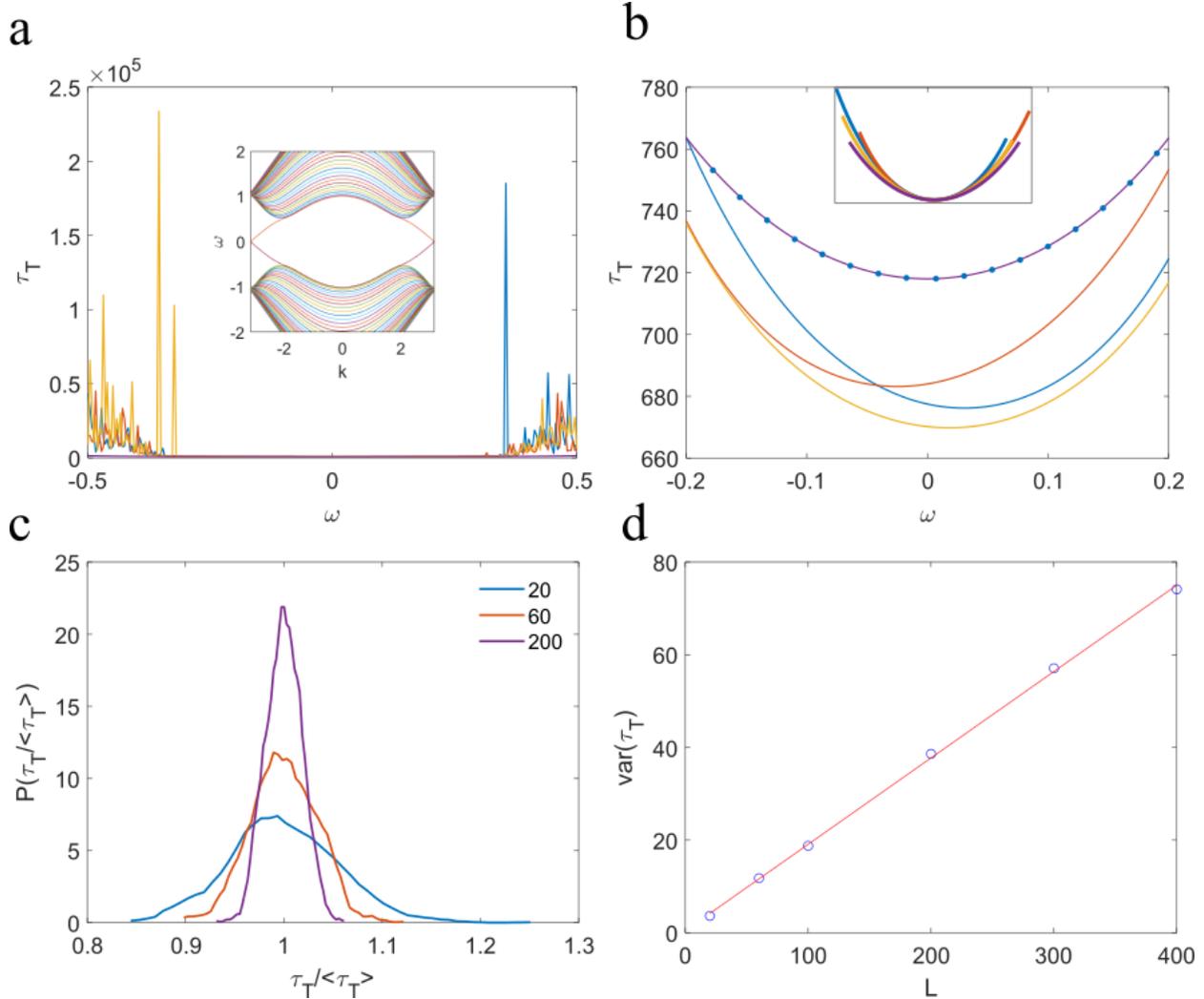

FIG. 5 (color online). Spectra and statistics of transmission time in band gap of disordered TI. Spectra of transmission time and in an ordered TI and in disordered samples with W=1.6, w=60, and L=400 over entire band gap (a) and in the center of the band gap (b). The dispersion relation in a periodic medium is shown in the inset in (a). Spectra of $\tau_T$ in the center of the band gap are displaced so that their minima coincide in the inset in (b). (c) Distribution of normalized transmission time and (d) the scaling of the variance of the transmission time.

The spectrum of each random configuration of a moderately disordered TI in the central region of the band gap shows no sharp structure, as seen in Fig. 5b since the contributions of the DOS of the edge mode which stretches across the band gap and of far-off-resonance modes in the passband vary slowly with frequency. However, fluctuations in the speed of the wave in different segments within a single sample results in different values of $\tau_T$ in different samples. As a result, the variations in $\tau_T$ within a single configuration over a narrow frequency range are smaller than the differences between configurations. Since only a fraction of the full range of $\tau_T$ found in the random ensemble is found in a single sample over a narrow range of frequency, disordered TIs are not ergodic.

The degree of correlation of the transmitted field with frequency when averaged over an ensemble of random systems relates spectral and temporal aspects of propagation in random media. In ergodic systems, the correlation function with frequency shift of the normalized transmitted field $E(\omega)$ is the Fourier transform of the ensemble average of the response to a delta-function pulse, $\langle |E(t)|^2 \rangle$ [47,48], known as the photon time-of-flight distribution. It is of interest to explore the extent to which the normalized field autocorrelation function, $C_E(\Delta\omega) = \langle E(0)E^*(\Delta\omega)\rangle / \langle |E(0)|\rangle\langle |E(\Delta\omega)|\rangle$, corresponds to the Fourier transform of the time-of-flight distribution in disordered TIs which are not ergodic. The field correlation function in a disordered ensemble of TIs of 500 samples of length $L$=400 with frequency shift $\Delta\omega$ relative to the frequency of the minimum of $\tau_T$ in the pristine TI of $\omega = 0$ is shown in Fig. 6a. The real part of $C_E(\Delta\omega)$ is shown as the solid blue curve, the imaginary part as the red dashed curve, and $|C_E|$, which gives the envelope of the oscillating functions, is shown as the solid yellow curve. Since $|E|$ is constant through the band gap, the decay within the band gap occurs because of variations between configurations in the phase shift of the fields between $\omega = 0$ and $\Delta\omega$. In a single disordered TI, the correlation function, shown in Fig. 6b, continues to oscillate with unit amplitude until the frequency reaches the passband where the transmitted amplitude falls.

The Fourier transform of $C_E$ in the disordered TI is shown as the blue curve in Fig. 6c. This cannot be precisely related to the ensemble average of the response to a delta-function pulse $\langle I_0(t) \rangle$ because the response of the system changes over the frequency range of the band gap. The variation of propagation variables with frequency is also a limitation in random media. Nonetheless, $C_E$ and the pulse response to a wide bandwidth pulse, $\langle I(t) \rangle$, have been demonstrated to be a Fourier transform pair to high accuracy in optical [47] and microwave [48] measurements in random media. The reason for this is that the correlation frequency, in which the field correlation falls appreciably and is approximately equal to the mode linewidth in random media, is small compared to the frequency range on which propagation variables change. In the case of a disordered TI near the band center, the frequency in which the magnitude of the field correlation function falls to 1/e of approximately 0.15 is not a negligible fraction of the band gap, but over this range the change in $\tau_T$ is small, as seen in Fig. 5b. We choose a pulse bandwidth of 0.2, which is larger than the correlation frequency but smaller than the width of the band gap, for the incident Gaussian pulse. It is not possible to choose a larger bandwidth because the pulse would then span the band gap over which the dispersion in $\tau_T$ is large enough that the transmitted pulse would be substantially broadened. With this choice, the Fourier transform of $C_E$, given as the blue curve in Fig. 6c is in reasonable agreement with of $\langle I(t) \rangle$.

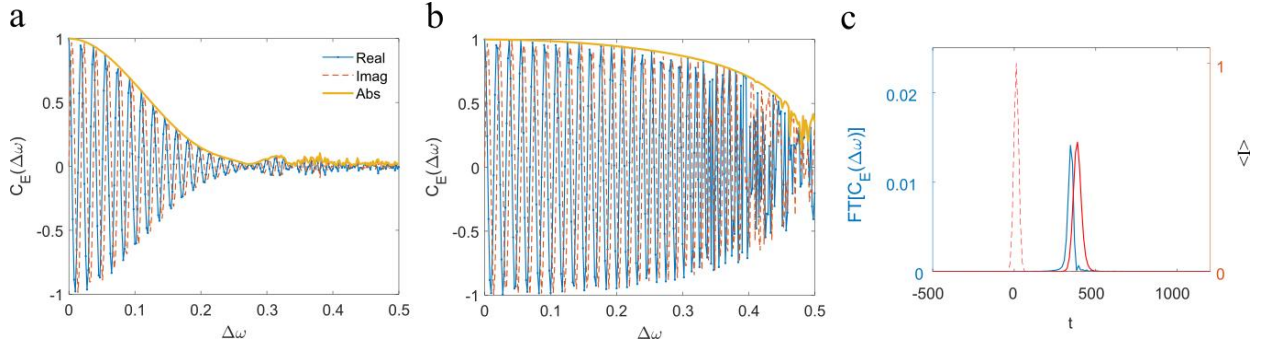

FIG. 6 (color online). Field correlation function and dynamics over random ensemble and single realization of disordered TIs. (a) Real and imaginary parts and the amplitude of field correlation function with frequency shift between the center of band gap at $\omega = 0$ and $\Delta\omega$ over random ensemble. (b) Field correlation function for a single random sample. (c) Comparison of the Fourier transform of $C_E$ (blue curve) and the response to a Gaussian pulse with bandwidth of 0.2 (solid red curve).

Incident Gaussian pulses (blue) with bandwidths of 0.2 and 0.6 peaked at the center of the band gap and the responses (red) in a single sample with W=1.6 and L=200 are shown in Figs. 7a,b, respectively. The incident 0.2-bandwith pulse and the response in a bent sample with disorder W=1.6 is shown in Fig. 7c. The energy distributions in a section of the pristine and disordered systems with a straight path for steady-state excitation at $\omega = 0$ are shown in Fig. 7d. The corresponding distribution for the bent path is shown in Fig. 7e. Because the energy density falls off so rapidly from the edge, the color bar encodes values of $|E|^{1/2}$. For the narrower linewidth incident pulse, the transmitted pulse appears to have the same shape as the incident pulse. For the broader band pulse, a long tail develops due to part of the spectrum that is closer to the band edge where $\tau_T$ is larger and the speed along the edge is smaller. There seems to be no discernable broadening of the pulse that encounters two bends, as seen in Fig. 7c. This demonstrates that pulse propagation near the band center in a single sample is robust in the presence of disorder and bending of the path. There is no evidence of an increase in the energy density at any point along the path. These results demonstrate that pulse propagation is robust in the band gap of a disordered TI even for ultrashort pulses with bandwidths which are a significant fraction of the width of the band gap.

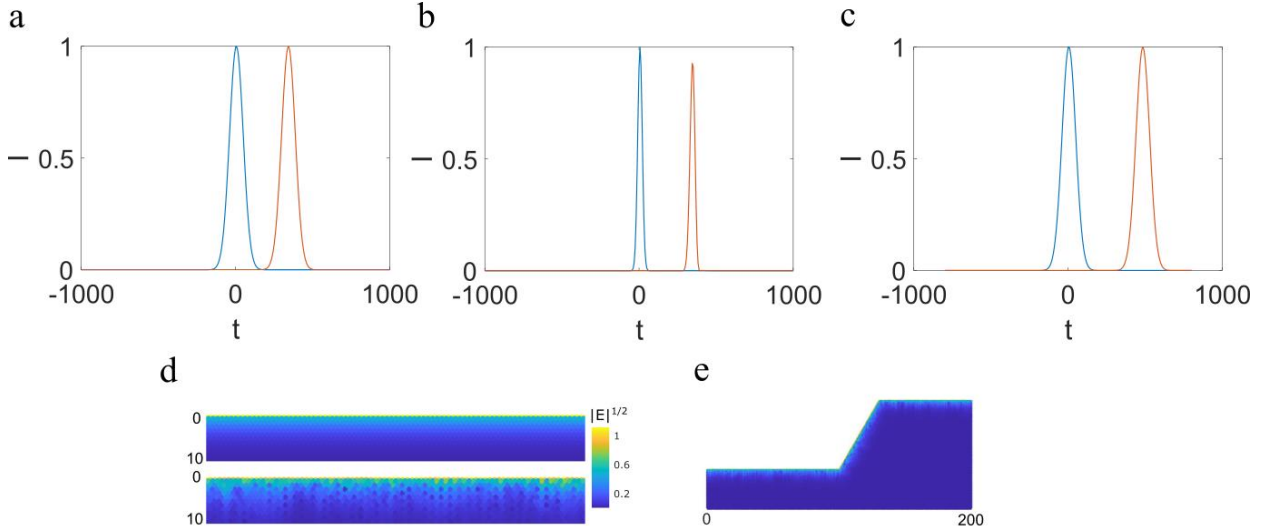

FIG. 7 (color online). Robust pulse propagation in the center of the band gap. (a-c) The blue and red curves show the time dependent flux in the incident Gaussian pulse and transmitted pulses, respectively. The bandwidth of the pulses is 0.2 in (a,c) and 0.6 in (b). The length of the straight path along the edge is 200 in (a,b). The pulse in the straight and bent paths in (a,c) are not significantly distorted in transmission through the TI. The wider-band pulse in (b) is slightly broadened by the dispersion in $\tau_T$ over the band gap, as can be seen in the reduction of the peak intensity of the transmitted pulse. The spatial distribution of $|E|^{1/2}$ is plotted in (d,e). The energy density along the path shows only short-range fluctuations.

### D. Propagation random one-dimensional nontopological media

We have seen that the statistics and scaling of $\tau_T$ in disordered TIs differ through the band gap with initial quadratic scaling where modes exist and linear scaling where they do not. Since these statistics provide a window on the nature of propagation in the medium, it is natural to compare these results to the statistics of propagation in trivial random 1D media. In random media, it is possible to explore the statistics of transmission as well as of the transmission time. The single parameter scaling theory of localization developed for random 1D media shows that the logarithm of the transmission coefficient, $\ln T$, self-averages with an average that scales linearly, $\langle \ln T \rangle = -L/\ell$ [49]. Further, the probability distribution of $\ln T$ approaches a Gaussian for $L \gg \ell$ with a variance of transmission of $2L/\ell$, giving, $\langle \ln T \rangle = -L/\ell \sim \text{var}(\ln T)/2$ [49,50]. Thus, $L/\ell$ is the universal dimensionless scaling parameter which determines both the centroid and variance of Gaussian distribution of $\ln T$ and so all statistical properties. However, because transmission is robust in disordered TIs, |E| and $T$ are constant, and $\ell$ is not a suitable transport parameter. The only transmission variable is the phase of the transmitted field. In perfectly nonreciprocal media, the transmission time, $d\varphi/d\omega = \tau_T$, the DOS, and energy excited in nonreciprocal media may be found in accord with Eq. (3). We have seen that the nature of propagation in disordered TIs can be described in terms of the statistics of $\tau_T$. The statistics of the transmission time have been measured in multichannel quasi-1D random media in the waveguide or wire geometry of constant cross section and reflecting sides [23] and in slabs [27], and have been calculated and compared to simulations for reflection from a medium connected to free space via a single channel [24], but the statistics of dynamics for transmission through a random 1D media have not been reported. These statistics are calculated using transfer matrix simulations in random binary layered samples and shown in Fig. 8. Samples are composed of layers with thickness selected randomly between 0 and 1 cm within a sample of length, $L$= 1m. The vacuum wavelength is 0.3 cm and the indices of refraction of the layers alternate between two values, $n_1$=1-$\Delta n$ and $n_2$=1+$\Delta n$. $\Delta n$ is selected to give the desired mean free path, which is determined via the relation $\langle \ln T \rangle = -L/\ell$. The statistics of $\tau_T$ are calculated for ensembles with $10^6$ samples.

The probability distribution $P(\tau_T/\langle \tau_T \rangle)$ in the random 1D ensemble described above is shown in Fig. 8a for different values of $L$ for $\ell = 0.5$ m. The statistics of transmission time differ sharply from those in disordered TIs both when modes are not present (Fig. 4c,e) and when they are (Fig. 4d,f). For $L/\ell < 1$, $P(\tau_T/\langle \tau_T \rangle)$ is narrow and peaked near unity since the wave is ballistic and is barely scattered. As $L$ increases, the peak of $P(\tau_T/\langle \tau_T \rangle)$ shifts to lower values of $\tau_T/\langle \tau_T \rangle$ and a long tail develops. In localized systems, the average linewidth of modes falls exponentially with thickness and the distribution of linewidths is wide since it depends upon the separation of the location center from the sample boundaries [33]. As a result, most frequencies fall outside the linewidth of the nearest mode and the DOS and $\tau_T$ for most frequencies are small. When the frequency falls near resonance with a localized mode, however, the DOS, and so $\tau_T$ is high. This results in a long tail in $P(\tau_T/\langle \tau_T \rangle)$ with decreasing weight as $L$ increases. The universality of statistics of $\tau_T/\langle \tau_T \rangle$ is demonstrated in Fig. 8b by the overlap of $P(\tau_T/\langle \tau_T \rangle)$ for different values of $L$ and $\ell$ but the same value of $L/\ell$. The scaling of the variance $\tau_T/\langle \tau_T \rangle$ is seen in Fig. 8c to be universal and given by $\text{var}(\tau_T/\langle \tau_T \rangle) = 0.090(L/\ell)^{1.88}$. This corresponds to $\text{var}(\tau_T) = (0.098/\ell^{1.88}v_+^2)L^{3.88}$, as seen in Fig. 8d. This nearly quartic scaling differs from the linear scaling in disordered TIs when modes are not present and the quadratic scaling when modes are present.

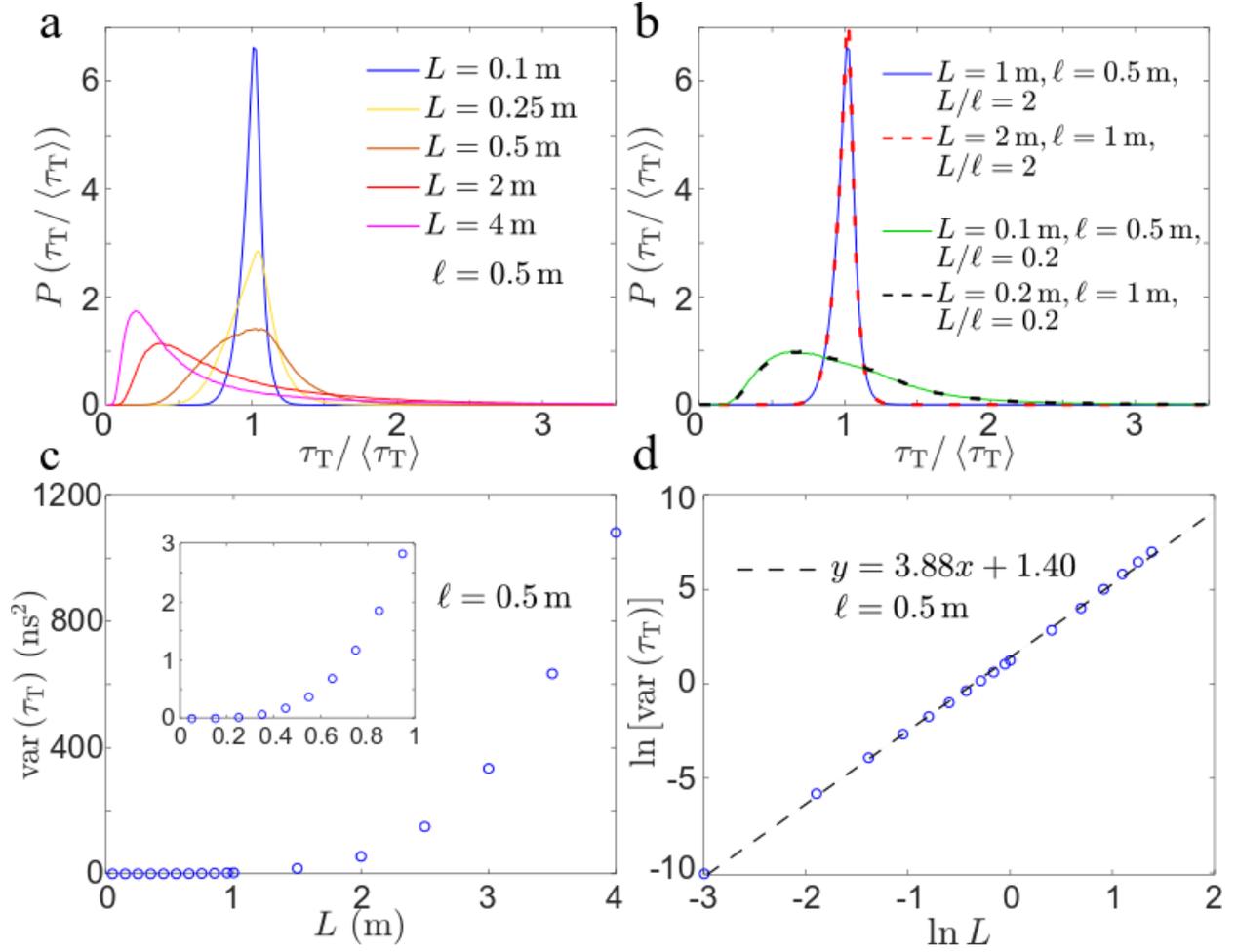

FIG. 8 (color online). Statistics of transmission time in 1D random media. (a) The probability distribution of transmission for different sample lengths with mean free path $\ell = 0.5$ m. (b) Comparison between the probability distribution of transmission time for different $L$ and $\ell$ but the same $L/\ell$ shows that statistics are universal an depend only on the parameter $L/\ell$. (c) The scaling of the variance of transmission time. (d) The same as (c) but in a log-log scale.

We note that $L/\ell$ is not a localization parameter in dimensions higher than one since the transmittance, which is the sum of transmission coefficients over all channels, $T = Tr(t^\dagger t) = \sum_1^N \tau_n$, also depends on the number of channels, $N$. Here, $t$ is the transmission matrix and the $\tau_n$ are the transmission eigenvalues. In ballistic or diffusive samples, in which transport is not significantly renormalized by coherent backscattering, the average transmittance is approximately given by $\langle T \rangle = N\ell/L$. The average transmittance is equivalent to the dimensionless conductance, $g$, which is the electronic conductance in units of the quantum of conductance, $e^2/h$. Thouless showed that, in diffusive samples, the dimensionless conductance is equivalent to the ratio of the average modal linewidth and the average frequency difference between modes, $\delta = g = \delta\omega/\Delta\omega$ [33,51]. The Thouless number, $\delta$, thus represents the degree of modal overlap. The inverse of the average mode spacing is the DOS, $\rho = 1/\Delta\omega$. When modes are weakly coupled to the boundaries at the two ends of the sample, and modes are narrower than the mode spacing, $\delta < 1$ and the modes are localized. Fluctuations of both steady state quantities, such as $T$, and dynamic quantities, such as $\tau_T$, are then large both within a single sample and between samples. On the other hand, when the modes are extended, they couple strongly to the boundaries. The modal linewidth is then larger than the spacing between modes, $\delta > 1$ and fluctuations are small. Unlike $T$, the Thouless number is not defined at each frequency in each random realization but is an average over an ensemble.

A dimensionless dynamical quantity which is defined at each frequency in each member of a random ensemble is $\tau_T/\langle\tau_T\rangle$. It is equal to the ratio of the DOS to its average, $\tau_T/\langle\tau_T\rangle = \rho/\langle\rho\rangle$, and, as we have seen, has universal statistics for standard random systems. Unlike $T$, whose average is a universal scaling parameter, the average of $\tau_T/\langle\tau_T\rangle$ is unity. However, the scaling of $\text{var}(\tau_T/\langle\tau_T\rangle)$ is universal and in 1D depends only on $L/\ell$, as does $\langle\ln T\rangle$ [24,27]. The statistics of $\tau_T/\langle\tau_T\rangle$ is not universal in disordered TIs, but, as we have seen, they provide a way of assessing the impact of modes of the passband on transport in disordered TIs.

### III. CONCLUSION

We have shown that the nature of propagation and the prospects for robust pulse transmission in disordered TIs in samples with different strengths of disorder and at frequencies throughout the band gap may be determined from the statistics of the transmission time. Since the transmission time in TIs is proportional to the DOS and the energy excited inside the medium, the transmission time and its statistics may be related to the quasi-normal modes of the medium and the continuum edge mode. Fluctuations of the central frequencies and linewidths of modes in the passband of the ordered TI due to disorder tend to raise $\tau_T$ near the band edge, while fluctuations in the edge mode tend to lower $\tau_T$ in the center of the band gap. In regions of the band gap in which modes are not introduced by disorder, $\tau_T$ self-averages so that $P(\tau_T)$ approaches a Gaussian distribution and $\text{var}(\tau_T)$ scales linearly. When modes are present in the band gap, $\text{var}(\tau_T)$ initially scales quadratically due to large fluctuations associated with narrow localized modes, which coexist with the extended edge mode. Beyond the length at which the full distribution of $\tau_T$ is sampled in a given sample, we expect that the variance will scale linearly.

Though there are significant fluctuations between samples near the center of the band gap, the spectrum of $\tau_T$ in a single configuration is smooth and even pulses short enough that their band width is a fraction of the band gap may be transmitted without significant broadening in samples of moderate length. In longer samples, pulses will be broadened by dispersion, as is the case in optical fibers. But the impact of dispersion can be minimized or compensated. The broadening of pulses by dispersion is minimized near the minimum of $\tau_T$ in the band gap and can be compensated by having successive metawaveguides with opposite dispersion. This can be achieved by having a waveguide with the frequency of the minimum in $\tau_T$ above the incident frequency in one section and below this frequency in the following section. Dispersion may also be compensated with use of chirped fiber Bragg gratings as is done in fiber optic communications [52]. Signals may be transmitted at several wavelengths in the band gap along a single metawaveguide as is done in wavelength division multiplexing in optical fiber telecommunication. Thus, high data rate data may be routed in metawaveguides as long as the band gap is not washed out by disorder [12,36]. In addition, disorder in nonreciprocal photonic TIs would not degrade the functioning of optical isolators for protecting pulsed laser sources.

In randomly disordered TIs, localized modes are formed near the band edge. But it is possible to create localized modes deep in the band gap adjacent to the edge of the TI which would coexist with the extended edge modes. This would enable narrow-line lasing in electronically or optically pumped defect states. Large area defects may be engineered so that the device can be pumped effectively while the linewidth of the mode may be controlled by the proximity of the defect mode to the edge of the TI. This TI laser is distinct from the recently demonstrated laser in which the gain is in the edge mode of the TI [53]. Our results show that the advantages of robust transmission are preserved for ultrafast signals in the central portion of the band gap for both linear and nonlinear devices.

## ACKNOWLEDGEMENTS

This work is supported by the National Science Foundation under EAGER Award No. 2022629 and by PSC-CUNY under Award No. 63822-00 51.

---------------------------

# Supplementary Material
# Dynamics of transmission in disordered topological insulators

Yuhao Kang, Yiming Huang and Azriel Z. Genack

Queens College and The Graduate Center of the City University of New York

Flushing, NY 11367, USA

## I. Relation between the transmission time and the intensity integral in a reciprocal random system

We simulate wave propagation in a reciprocal random system without loss or gain which is connected to the surroundings via single-channel leads on both sides of a sample, as shown in Fig. S1a. The transmission time, DOS, and sum of the integrals of intensity over the sample for unit flux flowing into the two leads are connected via the set of relations, $\tau_T = \pi\rho = U/2$ [1–3]. The equivalence of $\tau_T$ and $U/2$ is shown in Fig. S1b. The spectrum of $\tau_T$ shows contributions of isolated and overlapping modes. The inset in Fig. S1b shows the transmission in this system.

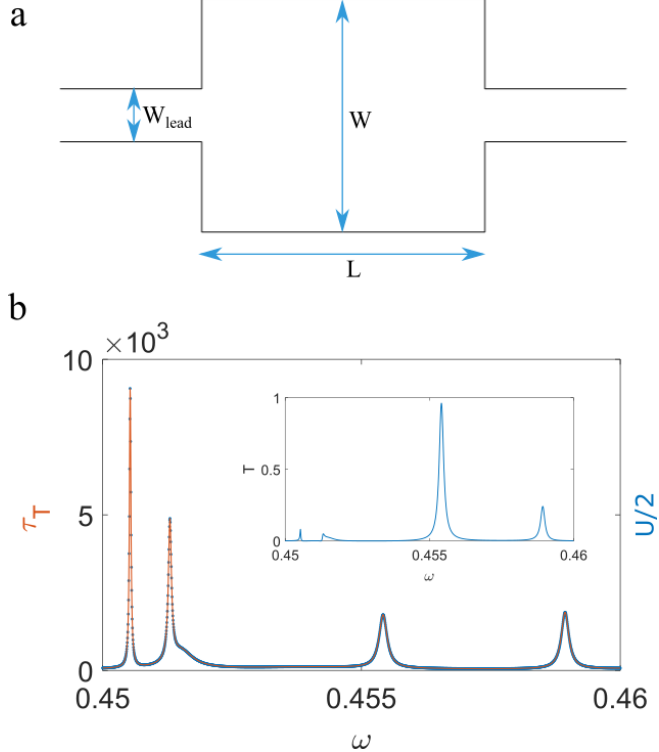

FIG. S1. Transmission time in a random medium. (a) The leads on the left and right support a single channel. The sample parameters are: $W_{lead}$=6, $W$=20, $L$=400. (b) The red curve is the transmission time. The blue dotted curve is one half the sum of the intensity integral for waves incident from the left and right leads, $U/2$. The inset shows the transmission spectrum.

## II. Relation between the Wigner time and the intensity integral in a topological insulator

We consider a TI following the Haldane model of the sample in Fig. 1a with disorder strength W=2, in which the waves in the upper and lower edges are coupled via localized modes in the bulk. The leads are a TI with the same structure as the pristine sample without disorder and support a mode moving to the right on the upper edge and a mode moving to the left on the lower edge. Because the system is neither perfectly reciprocal, with $\tau_T = \pi\rho = U/2$, nor nonreciprocal with unit transmission, with $\tau_T = 2\pi\rho = U$, the transmission time is not proportional to the DOS and to $U$, as can be seen in Fig. S2a. However, the Wigner time is proportional to the DOS and to $U$, $\tau_W = 2\pi\rho = U$. This is confirmed in the spectrum in Fig. S2b. The baseline in Fig. S2 is due to the edge mode.

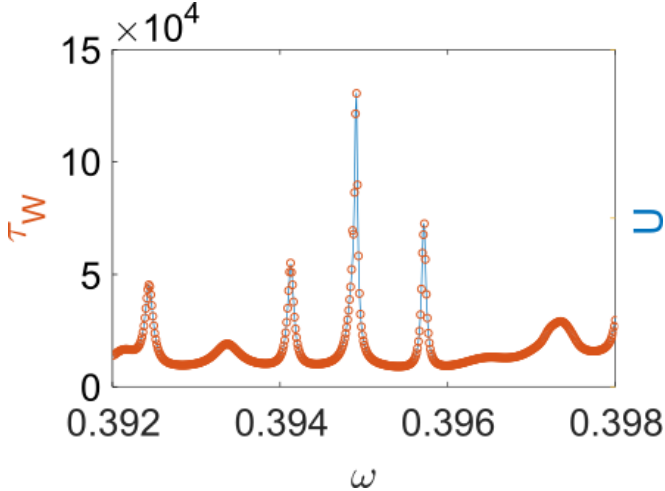

FIG. S2. Comparison of spectra of the transmission time, Wigner time and of excitation in a strongly disordered TI in which localized modes in the bulk couple the upper and lower edges of sample. The sample parameters are: W=2, w=60, and L=200. (a) The red circles are the transmission time and the blue curve is one half the sum of the integrals of intensity for the wave incident from both sides of the sample. (b) The red circles are the Wigner time and the blue curve is the sum of the integrals of intensity for the wave incident from both sides of the sample.

### III. Lorentzian decomposition of the transmission time

The transmission time and DOS a disordered TI without loss or gain are related as [4]

$$\frac{1}{2\pi}\tau_T = \rho(\omega) = \rho_0(\omega) + \frac{1}{\pi}\sum_n \frac{\Gamma_n/2}{(\omega-\omega_n)^2 + (\Gamma_n/2)^2} \quad \text{(S1)}$$

The first term in the expression for the DOS is the contribution of the edge mode, which varies slowly with frequency. The second term describes the contribution of narrow quasi-normal modes. We use Eq. (S1) to fit the transmission time for the spectrum shown in Fig. 3a to obtain the central frequencies and linewidths of the modes. The Lorentzian modes obtained are displayed in green in Fig. S3. The Lorentzian lines correspond to the contributions of individual modes to the transmission time. In this simulation, the resonant peak is $\approx 10^4$ and the background associated with $\rho_0$ is $\approx 10^2$.

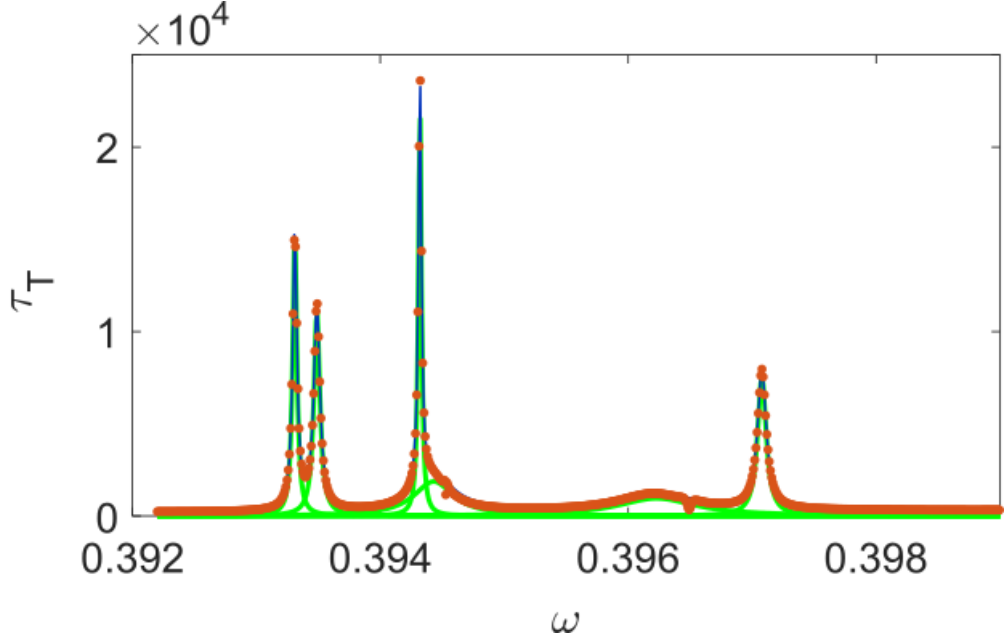

FIG. S3. Modal decomposition of the transmission time. The curve of red dots is the transmission time and the blue curve is the fit of Eq. (S1) to the time delay. The green curves are the contribution of individual modes. There is a slowly varying background of ≈200 due to the edge channel and modes outside of this frequency range. The dip at $\omega=0.3965$ at which the fit fails is due to reflection at the lower edge.

## IV. Measurement of the probability density of the transmission time in a photonic TI

We measured the probability density of the transmission time of microwave radiation along the edge between a quantum-spin-Hall (QSH) and quantum-valley-Hall (QVH) photonic crystal. The boundary between the QSH and QVH crystals, with Chern numbers of 1and 0, respectively, supports a single edge mode. The parameters of the sample shown in Fig. S4a are given in Ref. [5]. Disorder is QSH crystal is introduced by pushing rods with a collar to randomly selected positions of the collar between two copper plates. Measurements are made in 15 random configurations. The edge mode is supported along the boundary between the QSH and QVH crystals. The bandgap for the pristine system is between 20.5 and 22 GHz. Figure S4b shows the spectrum of transmission time for different disorder strengths. Inside the bandgap, the transmission time for the edge mode to pass through the 40-cm-long boundary line is approximate 4 ns. At the band edges, the transmission time increases due to the excitation of modes in the bulk. The probability distribution of transmission time between 20.8 and 21.5 GHz when 20 rods are disturbed randomly is shown in Fig. S4c. The distribution is asymmetrical with the peak near the ballistic time in the TI without disorder. The probability distribution is for a frequency range that includes the central region of the band gap as well as a portion closer to the band edge. The separate contributions of these regions would contribute, a Gaussian component and a long-time tail to the distribution, as in Figs. 5c and 4d of the main text, respectively.

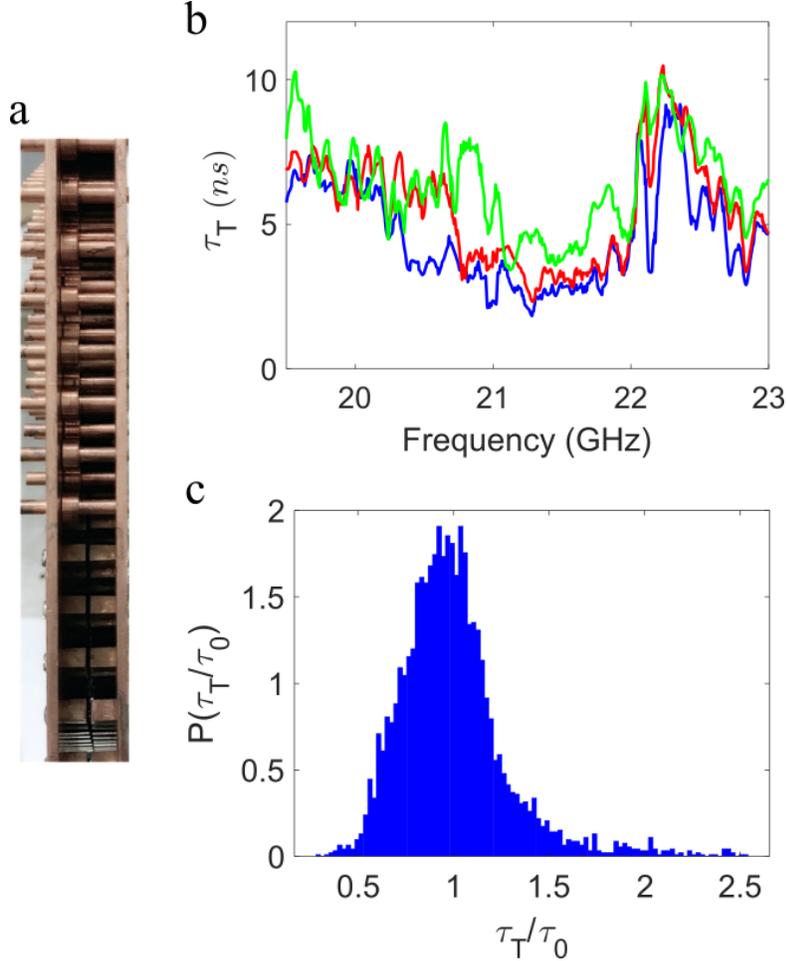

FIG. S4. Measurement of transmission time. (a) Photo of random configuration of the TI is Fig. 1a of Ref. [5]. (b) Plots of transmission spectra in a pristine system (blue curve) and in systems in which 10 rods (red curve) and 20 rods (green curve) are displaced. (c) Probability distribution of the transmission time within the bandgap between 20.8 and 21.5 GHz.

**V. Correlation of integrated intensity in two halves of a TI**

We seek to understand the source of the quadratic scaling of the distribution of the transmission time in the system with moderate disorder, as seen in Fig. 4f, in samples in which backscattering is prohibited. We show in the scatterplot of Fig. S5a in samples of total length $L=400$ that the integrals of the intensity within the two halves of random samples, $U_{1(2)}$, are uncorrelated. There are occasional large fluctuations in $U_1$ and $U_2$, but these tend not to occur in both halves of the same sample. Only in one of 400 samples is there a large fluctuation in both halves of the sample, as seen in Fig. S5.

The bulk of the distributions of both $U_1$ and $U_2$ and their sum $U$ normalized to the respective averages coincide and they all exhibit long tails. This corresponds to a variance of the bulk of the distribution of integrated intensity that scales quadratically, in accord with Figs. 4d,f. When the full range of values of $U/\langle U \rangle$, and so of $\tau_T/\langle \tau_T \rangle$ are included in the calculation of the variance, the result depends on just a few large values, which can be larger than 100, so that the variance cannot be determined based on the ensemble of 1,000 samples studied in this work. The large values of intensity and delay time arise on resonance with narrow modes that are distant from the edge.

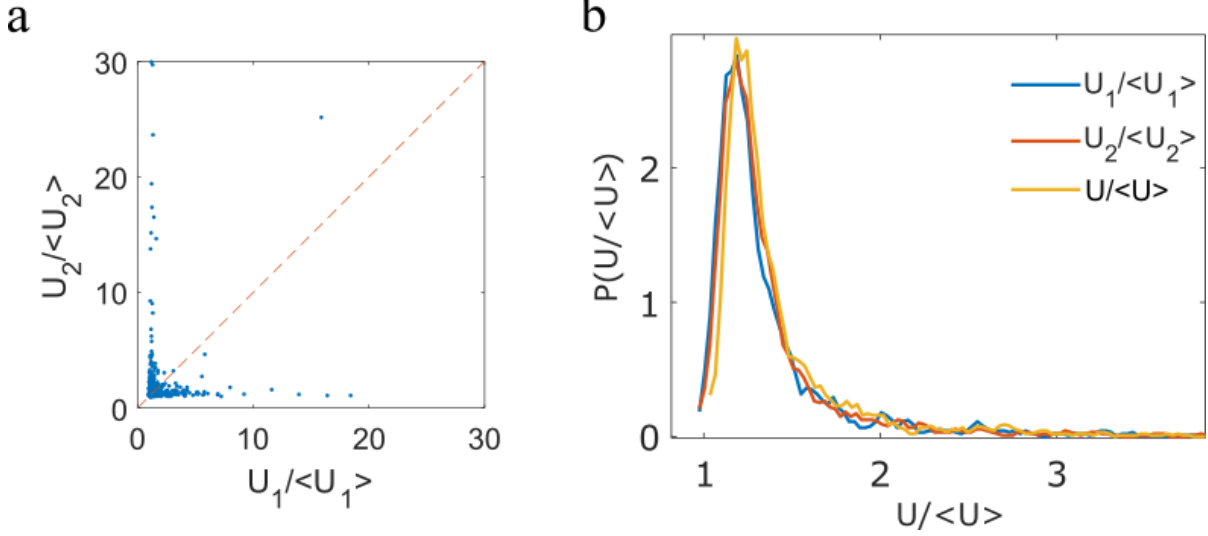

FIG. S5. Integral of intensity in a TI system with W=1.6, Width=60 and length=400. (a) The uncorrelated relation between the intensity integral of the first and second half part of the system. (b) The long-tail distribution of $I_{1(2)}$ and I normalized by their average.

**VI. Statistics of transmission time in nearly periodic topologically trivial 1D media**

Transfer matrix simulations of the transmission time carried out in a periodic and in three disordered photonic crystals are shown in Fig. S6. Sample of length $L=0.80$ m are composed of alternating layers of indices of refraction of 1 and 2. In the periodic system, the layers all have the same thickness of $d=0.02$ m, while the thickness of each layer is chosen from a rectangular distribution with thickness in the range $[0.95, 1.05]d$ in the random samples. Spectra of $\tau_T$ in the center of the band gap are shown in Fig. S6b. $\tau_T$ is in the disordered system is generally higher than in the periodic photonic crystal (purple curve). The probability distribution of $\tau_T$ normalized by its average, $P(\tau_T/\langle\tau_T\rangle)$, is shown in Fig. S6c for three lengths. For each length $\langle\tau_T\rangle$ is greater than the value in the periodic system, which is expected since the disorder enhances the wave transport in the bandgap and thus increase the DOS. The same might be expected for the disordered TI. But the contribution of the modes of the passband to the DOS in the center of the band is small and so the drop of $\langle\tau_T\rangle$ at the center of the band of 5% seen in Fig. 5 is largely the result of a drop in $\tau_T$ for the edge mode. Because the local DOS (LDOS) decays with depth into the sample in the band gap of trivial systems, $P(\tau_T/\langle\tau_T\rangle)$ and $\text{var}(\tau_T)$ saturate with length, as seen in Fig. S6c,d.

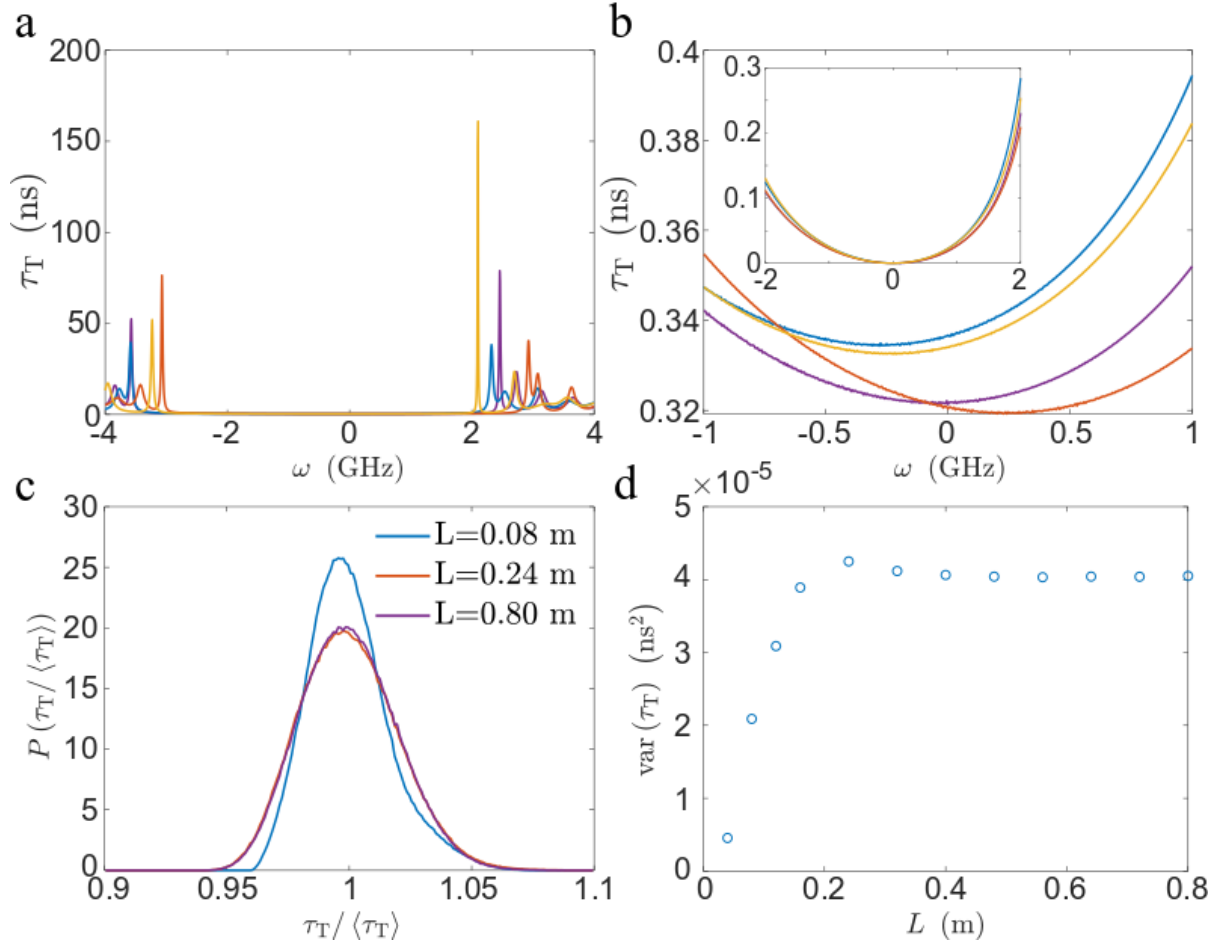

FIG. S6 (color online). Spectra and statistics of transmission time in the band gap of periodic and perturbed 1D nontopological media. Spectra over the entire band gap, (a), and in the center of the band gap, (b). (c) Distribution of normalized transmission time for different lengths, and (d) the scaling of the variance of the transmission time.

**VII. Scaling of the logarithm of transmission time in a trivial 1D system**

We have seen in Fig. 9 that the statistics of $\tau_T/\langle\tau_T\rangle$ in 1D which are the same as the statistics of the normalized DOS, $\rho/\langle\rho\rangle$, are universal with $\text{var}(\tau_T/\langle\tau_T\rangle) = 0.090(L/\ell)^{1.88}$. Another measure of universal fluctuations in the DOS is the average of the logarithm of the normalized transmission time $\langle\ln(\tau_T/\langle\tau_T\rangle)\rangle$. For $L \gg \ell$, $\langle\ln(\tau_T/\langle\tau_T\rangle)\rangle$ falls linearly and approaches a slope of approximately 0.09, as seen in Fig. S6.

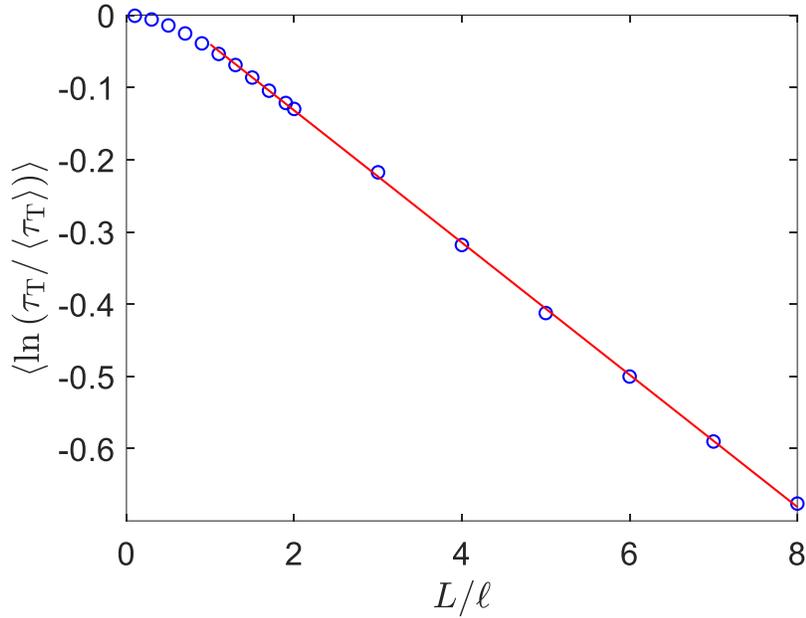

FIG. S7. Scaling of the average of the logarithm of normalized transmission time in nontopological trivial 1D random media.